\documentclass[aip,preprint]{revtex4-1}

\usepackage{graphicx}
\usepackage{dcolumn}
\usepackage{bm}
\usepackage{amssymb}
\usepackage[T1]{fontenc}
\usepackage{textcomp}
\usepackage{color}
\newcommand{\units}[3][]{$#1\mathrm{#2\,#3}$}
\newcommand{\mum}{\mbox{\textmu m}}

\begin{document}

\title{Experimental observation of the interaction of propagating spin waves with N\'eel domain walls in a Landau domain structure}

\author{P. Pirro}
\altaffiliation{currently: Institut Jean Lamour, Universit\'e Lorraine, CNRS, 54506 Vandoeuvre-l\`es-Nancy, France}
\affiliation{Fachbereich Physik and Landesforschungszentrum OPTIMAS, Technische Universit\"at
Kaiserslautern, 67663 Kaiserslautern, Germany}
\author{T. Koyama}
\affiliation{Department of Applied Physics, Faculty of Engineering, The University of Tokyo, Bunkyo, Tokyo 113-8656, Japan}
\author{T. Br\"acher}
\affiliation{Fachbereich Physik and Landesforschungszentrum OPTIMAS, Technische Universit\"at
Kaiserslautern, 67663 Kaiserslautern, Germany}
\affiliation{Graduate School Materials Science in Mainz, Gottlieb-Daimler-Strasse 47, 67663 Kaiserslautern, Germany}
\author{T.~Sebastian}
\altaffiliation{currently: Helmholtz-Zentrum Dresden - Rossendorf, Institute of Ion Beam Physics and
Materials Research, Bautzner Landstra\ss{}e 400, 01328 Dresden, Germany}
\affiliation{Fachbereich Physik and Landesforschungszentrum OPTIMAS, Technische Universit\"at
Kaiserslautern, 67663 Kaiserslautern, Germany}
\author{B. Leven}
\affiliation{Fachbereich Physik and Landesforschungszentrum OPTIMAS, Technische Universit\"at
Kaiserslautern, 67663 Kaiserslautern, Germany}
\author{B. Hillebrands}
\affiliation{Fachbereich Physik and Landesforschungszentrum OPTIMAS, Technische Universit\"at
Kaiserslautern, 67663 Kaiserslautern, Germany}

\begin{abstract}
The interaction of propagating dipolar spin waves with magnetic domain walls is investigated in square-shaped microstructures patterned from the Heusler compound Co$_2$Mn$_{0.6}$Fe$_{0.4}$Si. Using magnetic force microscopy, the reversible preparation of a Landau state with four magnetic domains separated by N\'eel domain walls is confirmed. A local spin-wave excitation using a microstructured antenna is realized in one of the domains. It is shown by Brillouin light scattering microscopy (BLS) that the domain structure in the remanence state has a strong influence on the spin-wave excitation and propagation. The domain walls strongly reflect the spin waves and can be used as spin-wave reflectors. A comparison with micromagnetic simulations shows that the strong reflection is due to the long-range dipolar interaction which has important implications for the use of these spin waves for excerting an all-magnonic spin-transfer torque.

\end{abstract}


\maketitle

The concept of magnon spintronics, i.e., the transport and manipulation of pure spin currents in the form of spin-wave quanta, the magnons, has attracted growing interest in the recent years. \cite{Serga2010,Lenk2011,Roadmap2014,SK_Kim2010,Kruglyak2010} Considering possible applications, like, i.e., spin-wave based logic and signal processing,\cite{Hertel2004,Pirro2011,Klingler_2014,Braecher2014} the local manipulation of spin waves in microstructured magnonic circuits is of paramount importance. 
To construct a device which, once configured, manipulates spin waves without energy consumption, magnetic domain walls are considered: Analytical and numerical studies predict that domain walls can be used to refract and reflect propagating spin waves \cite{Macke2010,SK_Kim2008} and to shift their phase. \cite{Hertel2004,Bayer2005} In addition, a scheme of moving domain walls by an all-magnonic spin-transfer torque has been proposed which makes use of the angular momentum transferred by spin waves which pass a domain wall. \cite{Mikhailov1984,Han2009,Yan2011,JS_Kim2012}

Several experimental studies have investigated the spin-wave spectrum in the presence of domain walls using a global (spatially homogenous) microwave or thermal excitation.\cite{Perzelmaier2005,Sandweg2008,Bailleul2007} However, despite the large quantity of numerical and theoretical studies, \cite{Hertel2004,Bayer2005,Macke2010,SK_Kim2008,Mikhailov1984,Han2009,Yan2011,JS_Kim2012,Roy2010,Wieser2010} no experiments concerning the interaction of locally excited, propagating spin waves with domain walls have been reported up to now. 

 \begin{figure}[b]
\begin{center}
\scalebox{1}{\includegraphics[width=0.75 \textwidth, clip]{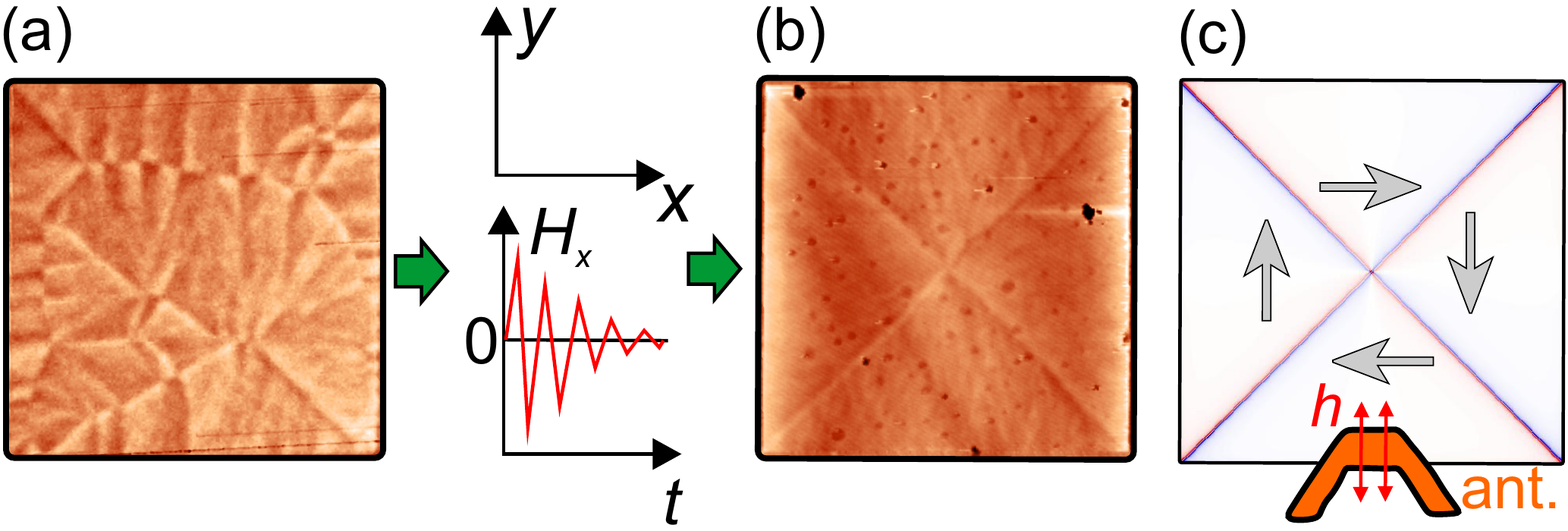}}
\end{center}
\caption{\label{Fig_Schema}(color online) (a),(b) MFM measurements of the remanence state of a CMFS square (edge length 20 $\mu$m) with the scheme of the field sequence to nucleate the Landau state. (a) Multi domain state after fabrication. (b) Landau state after external field sequence. (c) Scheme of the structure with microstructured antenna to excite spin waves. The red-blue color code shows the simulated divergence of the magnetization.}
\end{figure}

In this Letter, it is demonstrated that the interaction of propagating spin waves with N\'eel domain walls can be explored in square shaped Co$_2$Mn$_{0.6}$Fe$_{0.4}$Si (CMFS) microstructures. Using microfocus Brillouin light scattering (BLS), the strong influence of the hysteresis-dependent remanence state on the spin-wave excitation and propagation is proven. For the probed dipolar dominated spin waves, a strong reflection at the domain wall position is found, in agreement with recent micromagnetic simulations.\cite{Macke2010} The domain walls and consequently, the reflection positions of the spin waves, can be controlled using small external magnetic fields. Thus, these N\'eel domain walls can be employed as tunable spin-wave reflectors and blockers in microstructured magnonic circuits. 

The used CMFS film of 30 nm thickness is sputtered on a MgO substrate with a 40~nm Cr buffer layer and a 3~nm Ta capping layer. The investigated square structure with an edge length of 20 $\mu$m is patterned subsequently from the film by electron beam lithography and argon ion milling. CMFS has been chosen for this study due to its low Gilbert damping and high saturation magnetization \cite{Oogane2010,Kubota2009,Liu2009} which results in an efficient spin-wave excitation and a long spin-wave decay length.\cite{Sebastian2012,Sebastian2013,Pirro2014} These properties are indispensable to separate the point of the spin-wave excitation from the point where the interaction with the domain wall takes place.

First, a characterization of the magnetic domain structure using magnetic force microscopy (MFM) is performed (see Fig.~\ref{Fig_Schema}(a), (b)). The magnetization configuration found directly after fabrication is a multi domain state, as exemplarily shown in Fig.~\ref{Fig_Schema}(a), which involves a large number of domains and different domain wall types (e.g. cross tie domain walls \cite{Hubert-Schaefer} visible in the upper left). However, after a field sequence with decreasing amplitude (starting from 40 mT) and alternating sign, a Landau pattern with four symmetric domains and a vortex in the center can be nucleated as illustrated in Fig.~\ref{Fig_Schema}(b). As expected from the material parameters of CMFS and the film thickness, and confirmed by micromagnetic simulations (using MuMax2 \cite{mumax} with parameters \cite{parameter}), the domain walls are N\'eel type, i.e. the magnetic moments rotate in the sample plane. The result of the simulation is shown Fig.~\ref{Fig_Schema}(c), where the divergence of the magnetization, which indicates the position of the domain walls, is shown in a red-blue color.
Due to the comparably large size of the structure, the Landau domain state is transformed back into the multi-domain state if a field of 10 mT or more is applied and the magnetization is relaxed to remanence without an alternating field sequence. 

\begin{figure}[t]
\begin{center}
\scalebox{1}{\includegraphics[width=0.75 \textwidth, clip]{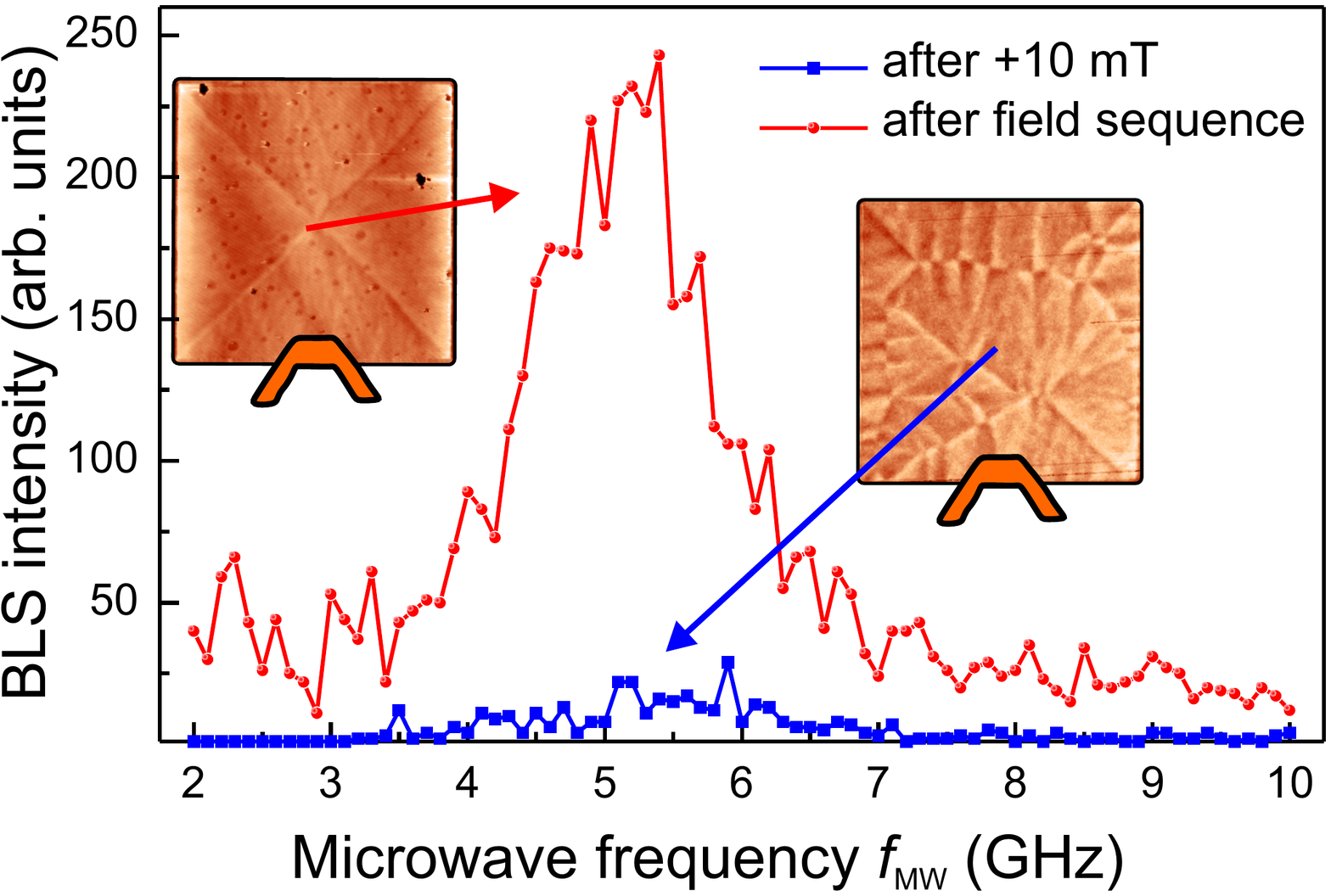}}
\end{center}
\caption{\label{Fig_Spectra}(color online) BLS spectra as a function of the excitation frequency $f_\mathrm{MW}$ recorded in remanence close to the center of the structure. Blue squares: spectrum in the multi-domain state. Red circles: excitation in the Landau domain state. In this case, due to the undisturbed propagation within the lower domain, a much higher spin-wave intensity is detected in comparison to the multi-domain state.}
\end{figure}

After the characterization of the domain structure, a microwave antenna (Cu, length in $x$: 3.8 $\mu$m, width in $y$: 1.4 $\mu$m, thickness in $z$: 300 nm) has been patterned on top of one of the edges of the square to locally excite coherent, propagating spin waves (see scheme in Fig.~\ref{Fig_Schema}(c), where the magnetization inside the domains of the Landau state is schematically illustrated by arrows). The waves are detected using microfocus Brillouin Light scattering (BLS),\cite{Sandweg2008,Sebastian2012,Sebastian2013,Pirro2011,Demidov2009,Braecher2014,Pirro2014} which allows to study the spin-wave intensity as a function of the magnetic field and the spin-wave frequency. In addition, it provides a spatial resolution better than \units{500}{nm}. 

Figure~\ref{Fig_Spectra} shows the spin-wave spectrum in remanence as a function of the excitation frequency, detected just below the center of the structure for the two previously described domain states. In the Landau state, the in-plane dynamic magnetic field of the antenna points perpendicularly to the local static magnetization, leading to a torque and, consequently, to an efficient excitation of spin waves. Due to the finite extensions of the antenna, this efficiency of the excitation is almost isotropic. However, as described in detail for CMFS thin film structures in Ref.~\cite{Sebastian2013}, the propagation of dipolar spin waves for in-plane magnetized films is strongly anisotropic and is restricted to an angle interval around the direction perpendicular\cite{angle_interval} to the static magnetization ($y$-direction for the lower domain in the Landau state).
The maximum group velocity of the waves is realized for a propagation perpendicular to the static magnetization,\cite{Kalinikos1986} i.e. towards the detection position indicated in Fig.~\ref{Fig_Spectra} by the end of the arrow. 
Thus, since no domain wall is present in-between the excitation source and the detection position, a clear spin-wave signal with a resonance around $f_\mathrm{MW} \approx 5.5$ GHz can be detected. This also shows that the exponential decay length of the spin wave, which is around 10 $\mum$ for CMFS films of the given thickness (compare Ref. \cite{Sebastian2012}), is long enough to propagate a sufficiently strong spin-wave signal to the vortex core and the domain walls to probe the interaction. In comparison, the spin-wave intensity in the multi-domain state is much lower. This indicates that the spin-wave excitation and propagation are strongly suppressed by the presence of the domain walls.

\begin{figure}[]
\begin{center}
\scalebox{1}{\includegraphics[width=0.75 \textwidth, clip]{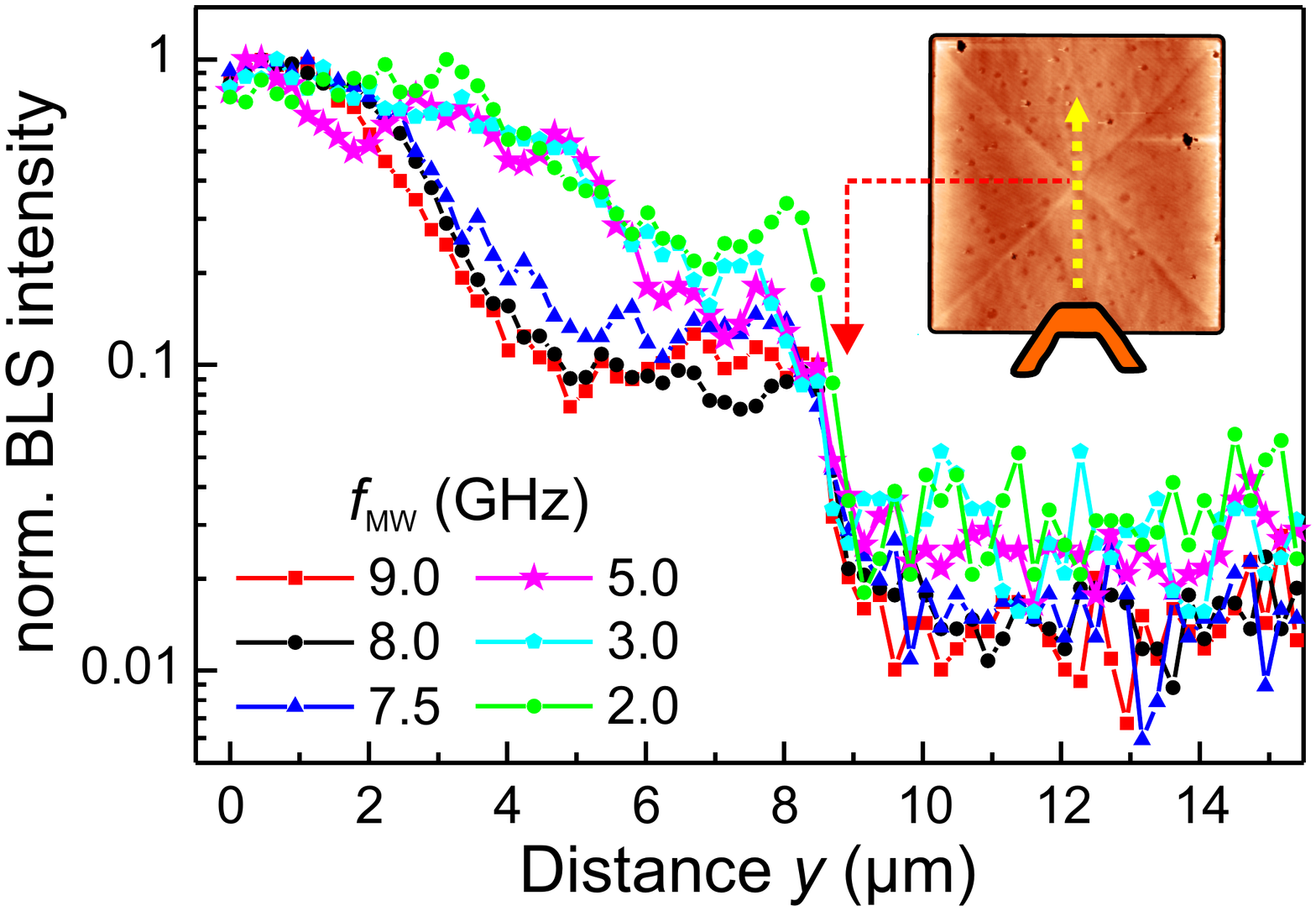}}
\end{center}
\caption{\label{Fig_diff_f} (color online) BLS intensity as a function of the distance from the antenna for different excitation frequencies $f_\mathrm{MW}$ ($\mu_0 H=0$ mT). At the position of the vortex core (around $y \approx$ 8 $\mu$m), a pronounced drop of the intensity is found for all spin-wave frequencies. The reflection of spin waves at this position leads to a standing wave pattern which causes the complex and non-exponential intensity decay between antenna and vortex core.}
\end{figure}

To directly investigate the interaction of the spin waves with the domain walls, the Landau state is used, since here, the position of the domain walls in remanence is well known and an efficient spin-wave excitation can be realized. 
Figure~\ref{Fig_diff_f} shows the individually normalized BLS intensity measured along the line through the vortex core indicated in the inset for different excitation frequencies $f_\mathrm{MW}$ at $\mu_0 H$=0 mT. Independent of the spin-wave frequency, a pronounced drop of the intensity is observed at the position of the vortex core. In contrast, only a spin-wave intensity close to the noise level of the measurement is detected behind the vortex core. Between antenna and vortex core, standing spin-wave patterns are formed, whose intensity distributions depend on the spin-wave frequencies. Due to the trigonal shape of the domain, these intensity patterns and the spin-wave reflection in general might also be influenced by the change of the effective quantization width for the spin waves propagating towords the vortex.

\begin{figure}[]
\begin{center}
\scalebox{1}{\includegraphics[width=0.5 \textwidth, clip]{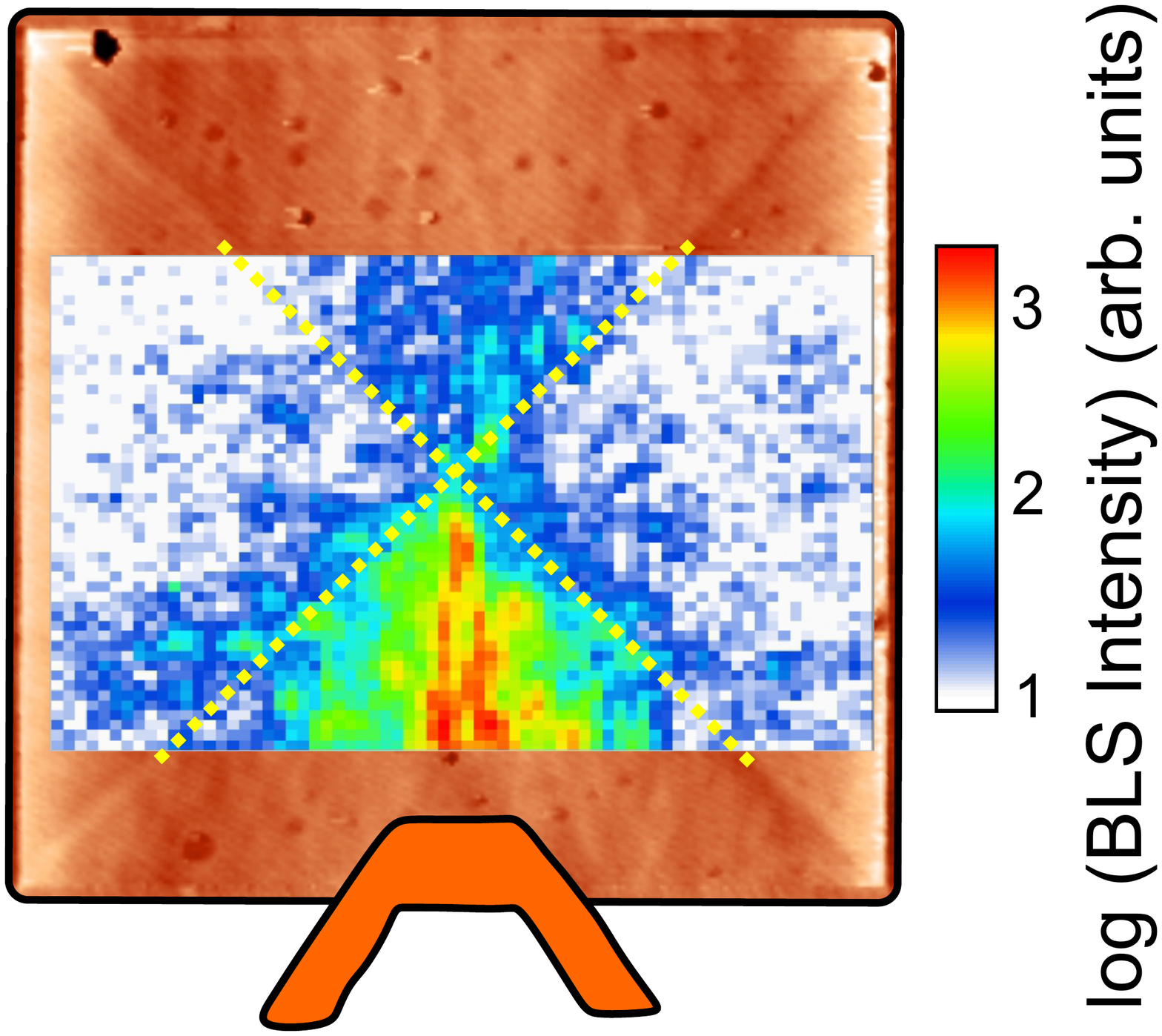}}
\end{center}
\caption{\label{Fig_2D_map}(color online) Two dimensional map (18x11 $\mu m^2$) of the color-coded BLS intensity (log scale) in remanence for an excitation frequency of $f_\mathrm{MW}= $4.55~GHz. The position of the domain walls is indicated by the dashed lines. Only a small part of the spin waves is transmitted through the domain walls from the lower domain into the adjacent domains.}
\end{figure}

 The spin-wave intensity is reduced also behind the N\'eel walls that separate two adjacent domains of the Landau state. This behaviour can be seen in Fig.~\ref{Fig_2D_map} which exemplarily presents a two dimensional BLS intensity map for a spin-wave excitation frequency of $f_\mathrm{MW}=4.55$ GHz at $\mu_0 H$=0 mT. Outside of the lower domain, where the spin-wave excitation takes place, a significant spin-wave intensity is found only at local spots close to the domain walls. Due to the absence of an external bias field, small defects in the structure or other inhomogeneities in the magnetic parameters can lead to considerable changes of the effective field. Thus, the existence of localized spin-wave modes close to the domain walls is probably caused by small deviations of the experimentally investigated structure from an ideal Landau domain state.

\begin{figure}[]
\begin{center}
\scalebox{1}{\includegraphics[width=0.75 \textwidth, clip]{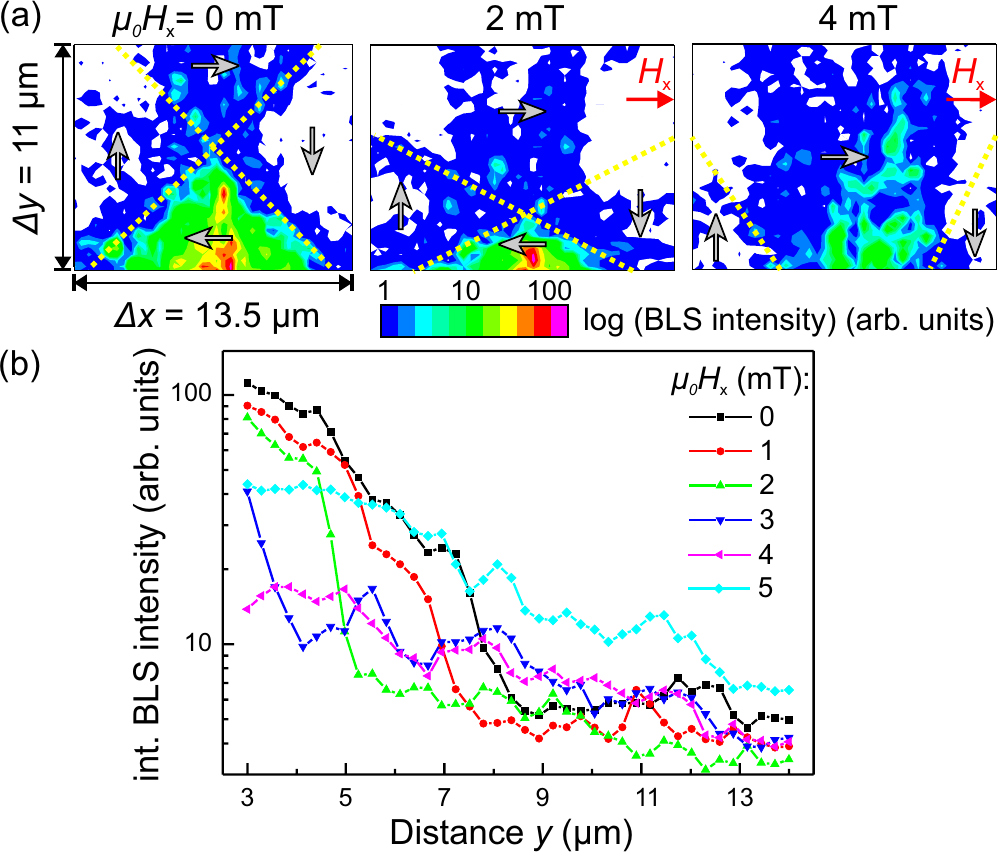}}
\end{center}
\caption{\label{Fig_field_shift}(color online) (a) Two-dimensional  BLS intensity maps (color coded, log scale) of the center of the structure for different external field $H_x$ ($f_\mathrm{MW}=$ 5.00~GHz). The domain walls and the directions of the magnetization are indicated schematically. (b) BLS intensity integrated over the $x$ coordinate. The shift of the position where the reflection occurs is clearly visible for fields up to $\mu_0 H_x=$ 3 mT. } 
\end{figure}

In general, one can conclude from Fig.~\ref{Fig_diff_f} and Fig.~\ref{Fig_2D_map} that dipolar spin waves excited in this experiment are strongly reflected by the N\'eel domain walls and the vortex core. These findings can be used to construct a spin-wave reflector whose position can be tuned by small external bias fields. To illustrate this principle, Fig.~\ref{Fig_field_shift}(a) exemplarily shows two-dimensional measurements of the spin-wave intensity for three different external bias fields $H_x$ ($f_\mathrm{MW}$ is fixed to 5.0 GHz). In Fig.~\ref{Fig_field_shift}(b), the intensity is shown integrated over the $x$ coordinate to illustrate the average intensity as a function of the distance $y$ to the antenna for different  $H_x$. 
With increasing $H_x$, the lower domain, which has a magnetization pointing opposite to the direction of the external field  $H_x$, shrinks, whereas the upper domain grows. Up to a field of $\mu_0 H_x$ = 3 mT, this leads to a shift of the position of the spin-wave reflection from $y \approx$ 8 $\mu$m to $y \approx$ 3 $\mu$m which is due to the displacement of the reflecting domain walls towards the antenna. For $\mu_0 H_x$ = 4 mT, the vortex core reaches the antenna and the spin-wave excitation is suppressed due to the inhomogeneous magnetization in the excitation area. For $\mu_0 H_x$ = 5 mT, the vortex core is pushed out to the edge of the structure. No abrupt drop of the intensity is observed anymore in this case. The reason for this is that no domain walls hinder the spin-wave propagation in $y$ direction anymore which leads to the increased spin-wave intensity at this field compared to lower field values. Thus, in the presence of domain walls, the propagation and spatial distribution characteristics of spin waves can be strongly altered using comparably low external fields. 

Further measurements (not shown) similar to the ones in Fig.~\ref{Fig_field_shift}, but with external fields applied at different in-plane angles show that the strong reflectivity is also present if the incident angle of the spin waves on the domain walls is changed.

It is interesting to compare the observed strong reflection of the dipolar dominated spin-waves at the N\'eel domain walls with the predictions of the systematic micromagnetic simulations in Ref.~\onlinecite{Macke2010}. There, it is predicted that the dipolar interaction in magnetic thin films leads to a strong spin-wave reflection for wavelengths which are larger than the N\'eel domain wall width $w$. In our case, $w$ is in the range of $\pi \sqrt{\frac{2 A}{\mu_0 M_s^2}} \approx$ 15 nm (see Ref. [\onlinecite{Hubert-Schaefer}]), whereas the minimal excited spin-wave wavelength, which is determined by the antenna size \cite{Demidov2009,Pirro2011}, is in the micrometer range. Thus, the high reflectivity found in the experiment is in agreement with the simulations and can be attributed to the long-range dipolar interaction. However, from the experimental results, it cannot be finally concluded if the domain wall itself or the different orientations of the two adjacent domains are responsible for the reflection. To answer this questions, further experiments using materials which show different domain wall configurations like, e.g., Bloch walls have to be performed.
Our observations can also have important implications for the use of the dipolar dominated waves in all-magnonic spin-transfer torque applications. The transfer of angular momentum from the spin wave to the domain wall occurs only if the spin wave is transmitted through the wall.\cite{Mikhailov1984,Yan2011} Thus, short wavelength, exchange-dominated spin waves, which can pass a domain wall without any reflection,\cite{Bayer2005,Yan2011} could be more suitable to exert an all-magnonic spin-transfer torque to a narrow magnetic domain wall. However, the so-called momentum-transfer effect observed in several micromagnetic simulations \cite{Moon2013} might allow for the movement of magnetic domain walls by dipolar spin waves since this effect relies on the change of the momentum of the waves due to reflection. This effect requires further studies with domain configurations with a lower stability than the Landau structure. 

To conclude, we have experimentally proven that magnetic domain walls have a significant impact on the excitation and propagation of dipolar spin waves. 
By the controlled nucleation of a Landau domain pattern, we were able to investigate the interaction of locally excited spin waves with individual domain walls. The measured spin-wave spectrum strongly depends on the hysteresis of the system which is mainly due to the pronounced reflection of the dipolar spin waves by the domain walls. Based on these findings, we demonstrated that domain walls can serve as flexible spin-wave reflectors whose position can be controlled by comparably small external fields.

\begin{acknowledgments}
Financial support by the Optimas Carl Zeiss doctoral program and the Graduate School Materials Science in Mainz is gratefully acknowledged. T. Koyama acknowledges the Alexander von Humboldt foundation for a postdoctoral fellowship. We gratefully acknowledge financial support by the DFG Research Unit 1464 and the Strategic Japanese-German Joint Research Program from JST: ASPIMATT.
\end{acknowledgments}


\begin{thebibliography}{99}
\bibitem{angle_interval}
For the used excitation frequencies in this publication, the angle of the spin-wave propagation is restricted $\pm$25$^{\circ}$ around the direction perpendicular to the static magnetization.
\end{thebibliography}
\end{document}